\documentclass {iaus}

\usepackage{amsmath}
\usepackage{amsfonts}
\usepackage{amssymb}
\usepackage{latexsym}
\usepackage[utf8]{inputenc}
\usepackage[english]{babel}
\usepackage{graphicx}

\title{Sampling effects in the emission line spectra of HII regions}
\author[Villaverde, Luridiana \& Cervi\~{n}o]{M. Villaverde, V. Luridiana, M. Cervi\~{n}o}
\affiliation{Instituto de Astrofísica de Andalucía (CSIC), Apdo. 3004, Granada 18080, Spain}

\pubyear{2006}
\volume{241}  %% insert here IAU Symposium No.
\date{?? and in revised form ??}
\jname{Proceedings Title IAU Symposium}
\editors{A.C. Editor, B.D. Editor \& C.E. Editor, eds.}
\begin{document}

\maketitle

\begin{abstract}
The combination of stellar population synthesis models and photoionization models allows a better understanding of the spectral features of HII regions and HII galaxies. In this work we show that sampling effects in the initial mass function (IMF) are very important in the low cluster mass case. To this aim, we compute photoionization models ionized by realistic clusters made up of various combinations of individual stars and clusters made up with a synthesis model. We discuss the differences in the position on diagnostic diagrams and their implications.
\end{abstract}

\firstsection

\section{Introduction}
Clusters modeled by stellar population synthesis models have been used as photoionization sources in several works (\cite{CM-H94}, \cite{Letal99}, \cite{SI03} among others).
Most of the current models assume that the Initial Mass Function (IMF) is completely sampled, but when the number of stars is small this assumption is incorrect. Therefore standard synthesis models cannot be applied to low mass clusters.
\cite{Cetal03}, \cite{CL04}, \cite{CL06} showed that this question can be treated from a probabilistic point of view in which the IMF is interpreted as  the probability distribution function of the stellar mass rather than its deterministic expression. Implementing the formalism developed by the mentioned authors, the sampling problem can be studied in a better way.
The aim of this work is to show the importance of sampling effects in the IMF for low mass clusters, the most abundant ones, and to emphasize the need of a probabilistic approach to this problem.

\section{Method and results}
Using Emission Line Diagnostic Diagrams (ELDDs) we compare the emission-line spectra obtained with a synthesis model (\cite{Cetal02}) with others obtained with combinations of individual stars of 20, 25, 50 and 100 M$_{\odot}$. In each of these combinations the stars have the same mass so that sampling effects are important. With these models we can explore a three parameter space: mass, number of ionizing photons (Q(H$^{0}$)) and number of stars. We consider three cases: fixed mass equal to 100 M$_{\odot}$; fixed Q(H$^{0}$) equal to the one of a 100 M$_{\odot}$ star; and fixed number of stars equal to N=1. All the models have been computed for ZAMS.

Using CLOUDY (\cite{Fetal98}) we compute the emission of model nebulae with Orion abundances ionized by the clusters described above and calculate their representative points in ELDDs.

Figure 1 (left) show the differences in a selected ELDD between the predictions of the synthesis model cases and those of the rest of the cases. The synthesis model cases fall either far from any of the stellar cases  or very close to the 50 M$_{\odot}$ stellar case. Moreover, the points corresponding to the synthesis model cases closely follow the tendency of the 50 M$_{\odot}$ cases. This is expected because the ionizing continuum for the synthesis model case is very similar in shape to the one 50 M$_{\odot}$ star (Fig. 1, right).
In all the cases the same abundances are used, but there is a considerable dispersion in the position on  ELDDs due to the differences in the ionizing continua of the stars used. So, if a comparison with theoretical grids of HII regions based on usual synthesis models were done, it could be wrongly deduced that our theoretical cases span a wide range in metallicity and/or age.
\begin{figure}[t]
\centering
\includegraphics[scale=0.2871]{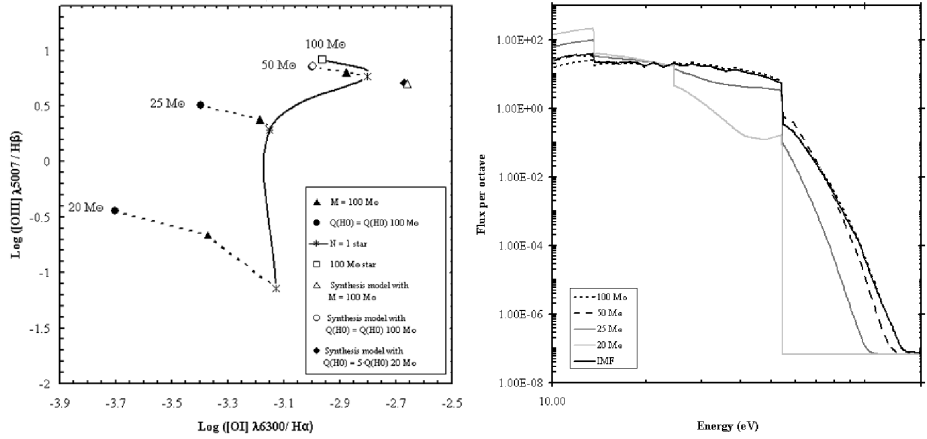}
\label{fig:1}
\caption{Left: [O III] $\lambda$5007/H$\beta$ vs. [O I] $\lambda$6300/H$\alpha$ intensity ratios, dashed lines join cases with stars of the same mass. Right: ionizing continua for fixed Q(H$^{0}$) cases.}
\end{figure}

\section{Conclusions}
The dispersion shown by empirical ELDDs can be attributed not only a spread in metallicity and age but also to IMF sampling effects, which introduce an additional degeneration. The smaller the cluster, the more severe the effect.
If we can infer from observations the value of Q(H$^{0}$) or the mass of the underlying ionizing cluster it's impossible to accurately determine, by means of a standard synthesis model, the corresponding point in none of the ELDDs given the separation between the models.
Therefore standard synthesis models should not be applied to low mass clusters because they would lead to misinterpretations of observational or theoretical results. In order to complement this work, the relative probabilities of the different cases will be computed in the next future using the probabilistic formulation followed by Cervi\~{n}o \& Luridiana (\cite{CL06}.
\linebreak
\linebreak
{\it Acknowledgements}. This work was supported by the Spanish project PNAYA2004-02703. MV is supported by a MEC-FPI fellowship, VL by a {\it CSIC-I3P}  fellowship and MC by a {\it Ram\'on y Cajal} fellowship.

\end{document}